\documentclass[sigconf,nonacm]{acmart}
\usepackage{multirow}
\usepackage{subcaption}

\AtBeginDocument{%
  }

\begin{document}

\title{Cluster-Level Experiments using Temporal Switchback Designs: Precision Gains in Pricing A/B Tests at LATAM Airlines}

\author{Nicolás Ferrari-Ortiz}
\affiliation{%
  \institution{LATAM Airlines}
  \city{Santiago}
  \country{Chile}}
\email{nicolas.ferrari@latam.com}

\author{Sebastián Orellana-Montini}
\affiliation{%
  \institution{LATAM Airlines}
  \city{Santiago}
  \country{Chile}}
\email{sebastian.orellanam@latam.com}

\author{Timur Abbiasov}
\affiliation{%
 \institution{ADC Consulting}
 \city{Amsterdam}
 \country{Netherlands}}

\author{Marie Garkavenko}
\affiliation{%
 \institution{ADC Consulting}
 \city{Amsterdam}
 \country{Netherlands}}
 \email{marie.g@adc-consulting.com}

\author{Rutger Lit}
\affiliation{%
 \institution{ADC Consulting}
 \city{Amsterdam}
 \country{Netherlands}}
 \email{rutger.l@adc-consulting.com}

\renewcommand{\shortauthors}{Ferrari-Ortiz et al.}

\begin{abstract}
  Experimentation is central to modern digital businesses, but many operational decisions cannot be randomized at the user level. In such a case cluster level experiments, where clusters are usually geographic come to rescue. However, such experiments often suffer from low power due to persistent cluster heterogeneity, strong seasonality, and autocorrelated outcome metrics, as well as common shocks that move many clusters simultaneously. On an example of airline pricing—where policies are typically applied at the route level and thus A/B test unit of analysis is a route—  we study \emph{switchback} designs to remedy these problems. In switchback designs each cluster (route in our case) alternates between treatment and control on a fixed schedule, creating within-route contrasts that mitigate time-invariant heterogeneity and reduce sensitivity to low-frequency noise. We provide a unified Two-Way Fixed Effects interpretation of switchback experiments that makes the identifying variation explicit after partialling out route and time effects, clarifying how switching cadence interacts with temporal dependence to determine precision. Empirically, we evaluate weekly and daily switchback cadences using calibrated synthetic regimes and operational airline data from ancillary pricing. In our evaluations, switchbacks decrease standard errors by up to 67\%, with daily switching yielding the largest gains over short horizons and weekly switching offering a strong and simpler-to-operationalize alternative.
\end{abstract}


\maketitle

\section{Introduction}

Modern digital businesses increasingly rely on experimentation to make product, pricing, and policy decisions with measurable causal impact. At large scale, even small improvements compound, motivating sustained investment in experimentation platforms and rigorous analysis pipelines \cite{kohavi2009controlled,kohavi2020trustworthy,xu2015infrastructure,bakshy2014planout}. When the standard assumptions of independent units and stable outcomes are approximately satisfied, user-level randomized controlled experiments can deliver high power and straightforward inference.

However, many operational settings do not permit user-level randomization. A treatment may operate at the level of a market, region, or shared resource (e.g., supply allocation, dispatch, system-wide pricing rules), or it may induce spillovers across users through competition, congestion, or social and marketplace interactions. In such cases, the classical ``no interference'' assumption is violated, and causal estimands and valid inference must explicitly account for interference structures \cite{hudgens2008interference,aronow2017interference}. Practitioners therefore often move to cluster-level randomization, most prominently in \emph{geo experiments} where non-overlapping geographic regions are assigned to treatment/control \cite{vaverkoehler2011geo,kerman2017tbr}.

Cluster-level experiments are conceptually simple but statistically challenging. The number of available (usually geographic) units is often small, regions are highly heterogeneous, and outcomes are strongly autocorrelated and seasonal, reducing effective sample size and inflating uncertainty \cite{kerman2017tbr,chen2022iroas}. These issues are especially acute when outcomes are driven by broad common shocks (e.g., macro demand changes, operational disruptions, or policy/regulatory shifts) that move many regions simultaneously, making counterfactual construction difficult even with sophisticated regression adjustments \cite{kerman2017tbr}.

Airlines pricing and ancillaries pricing in particular is a setting where clean user-level randomization is not feasible.
Prices are set at the route level, and existing commercial systems do not allow treatment assignment to be updated with the granularity or frequency that full randomization would require. As a result, the operationally viable alternative is to use parallel A/B tests in which entire routes are assigned to treatment or control. In practice, this design faces substantial limitations.

Routes differ persistently in passenger mix, booking patterns, and flight duration, generating structural heterogeneity that is difficult to absorb. In addition, demand exhibits strong serial dependence, reducing the effective number of independent observations.

To address this challenge, we evaluate \textbf{switchback} designs (further referred as \textbf{SB}) in which each route alternates between treatment and control on a fixed schedule. 
SBs offer a key advantage: they reshape the identifying variation relative to traditional route-level A/B tests. By alternating treatment states within the same route, they create within-unit contrasts that eliminate time-invariant heterogeneity, which would otherwise inflate sampling variability. At the same time, the regular switching cadence prevents short-term demand fluctuations from accumulating into persistent shocks, thereby weakening the serial dependence that complicates inference in temporal settings. As shown by \cite{bojinov2023design}, \cite{xiong2023}, and \cite{wen2025}, these features produce tighter confidence intervals without requiring strong assumptions about time-series dynamics.

Methodologically, we contribute by providing a unified Two Way Fixed Effects interpretation of SB designs that clarifies how temporal dependence and switching frequency affect statistical precision. 
By making explicit the identifying treatment variation after partialling out unit and time fixed effects, and analyzing how persistent shocks enter the estimator under different assignment schemes, we show how SB designs reduce exposure to low-frequency noise and increase effective sample size in the presence of autocorrelation. 
This perspective complements existing work on SBs by offering a transparent, variance-based explanation that links familiar econometric objects to concrete experimental design choices. To assess performance of SB designs, we adopt a comprehensive empirical strategy including analysis both on synthetic and real world data. Using LATAM’s operational data, we consider two cadences: weekly switches (every seven days) and daily switches (every calendar day). 
We show that in practice SBs lead to dramatically shorted run times under various scenarios with daily SB offering additional advantages at shorter experiment durations.

\section{Setup, Related Work, and Contribution}

\subsection{Setup}\label{sec:setup}

We consider a panel experiment with units indexed by $i = 1,\dots,N$ and time periods indexed by
$t = 1,\dots,T$. For each unit--time pair we observe an outcome $Y_{it}$ and a binary treatment
indicator $D_{it} \in \{0,1\}$. The outcome may exhibit temporal dependence due to slowly evolving
demand conditions, operational frictions, or other persistent shocks.

Treatment assignment varies over time according to a prespecified experimental design. 
We study designs that differ only in how treatment status evolves over time within units. 
In particular, we distinguish between \emph{fixed assignment} schemes, where each unit is assigned permanently to treatment or control, and \emph{SB designs}, where treatment alternates over time so that each unit experiences both treatment and control states during the experiment. 
Apart from this difference in assignment structure, all other aspects of the experimental environment are held fixed.

The estimand of interest is the average contemporaneous treatment effect, denoted by $\beta$.
Throughout the paper, outcomes are modeled using a two-way fixed effects (TWFE) specification,
\begin{equation}
Y_{it} = \beta D_{it} + \Gamma_i + \Theta_t + \epsilon_{it},
\label{eq:twfe_setup}
\end{equation}
where $\Gamma_i$ captures time-invariant unit heterogeneity, $\Theta_t$ captures common shocks affecting all units in period $t$, and $\epsilon_{it}$ is a residual error that captures residual variation in outcomes not explained by treatment assignment or fixed effects. 
We do not require $\epsilon_{it}$ to be independent or identically distributed across units or time.
The TWFE framework defines a common estimand and comparison basis across all assignment schemes considered.
We refer to \cite{wooldridge2021} for a detailed discussion of the interpretation and properties of the TWFE model.

Our focus is not on identification, which follows from randomization under each design, but on
how different assignment schemes affect the sampling variability of the estimator of $\beta$ in
the presence of temporal dependence. 
By holding the estimand, outcome model, and estimator fixed, we isolate the role of experimental design, specifically treatment timing and switching frequency, in determining statistical precision, effective sample size, and the runtime required to detect economically meaningful effects.

\subsection{Related Work}
Our analysis connects to several areas of existing literature.

First, it connects to work on \emph{SB and temporal experimental designs}. 
Recent papers formalize SB experiments, study bias from carryover effects, and analyze inference under temporal dependence. 
The study conducted by \cite{bojinov2023design} provides a foundational treatment of SB design and analysis.
They analyze identification and bias in the presence of temporal interference, and show how design choices such as block length and switching frequency affect consistency and the validity of inference. Their focus is on ensuring unbiased and well-defined estimation under carryover, rather than on how switching frequency influences variance or statistical precision.
Subsequent work studies statistical properties under dependence and interference, including \cite{hu2022switchback}, \cite{xiong23a}, and \cite{xiong2024data}, who analyze bias--variance tradeoffs and design choices in temporal experiments. 
This literature emphasizes consistency, bias, and asymptotic rates, often under explicit mixing or carryover assumptions.

Second, our work relates to the time-series literature on variance inflation under autocorrelation.
It is well known that positive serial correlation increases the variance of sample averages and
reduces effective sample size, as captured by long-run variance formulas and heteroskedasticity- and
autocorrelation-consistent (HAC) covariance estimators (e.g., \cite{hannan1970multiple};
\cite{neweywest1987}; \cite{andrews1991heteroskedasticity}; \cite{hamilton1994time};
\cite{shumwaystoffer2017}). 
Differencing and related transformations are classical tools for reducing low-frequency dependence in such settings.
While the term SB experiment is relatively recent, the underlying idea has clear roots in time-series analysis, where autocorrelation is the central challenge. Early work on intervention analysis studied repeated and alternating interventions in autocorrelated systems as a way to separate treatment effects from trends and persistence (\cite{boxjenkins1970}; \cite{boxtiao1975}). 
More recently, these ideas have re-emerged in applied experimentation settings under the label of SB designs, particularly in online platforms and operational experiments, where time-based alternation is used to mitigate serial dependence through within-unit comparisons.

Third, our analysis builds on the panel data literature on two-way fixed effects (TWFE) estimation.
The TWFE framework makes explicit how identification and precision depend on within-unit and
within-period variation after partialling out unit and time fixed effects (e.g., \cite{wooldridge2010}; \cite{angristpischke2009}; \cite{wooldridge2021}). 
TWFE estimators are widely used in experimental and quasi-experimental settings, including difference-in-differences designs, but their interaction with temporal experimental designs and serial correlation has received less explicit attention.

Related extensions study SB designs under spatial interference and dynamic decision-making (\cite{jia2023clustered}, \cite{wen2024switchback_rl}), which address complementary settings beyond the panel experiment framework considered here.

\section{Methodology and Design}

\subsection{Estimation Model}

Building on the experimental setup introduced above, we formalize the estimation framework used to compare statistical precision across assignment schemes. 
We estimate the model introduced in Section~\ref{sec:setup}, reproduced here for convenience.
Throughout the analysis, treatment effects are estimated using a two-way fixed effects (TWFE) specification, which is held fixed across all designs to isolate the role of assignment structure.

Specifically, outcomes are modeled as
\begin{equation}
Y_{it} = \beta D_{it} + \Gamma_i + \Theta_t + \epsilon_{it},
\label{eq:twfe}
\end{equation}
where $i$ indexes units and $t$ indexes time periods. 
The parameter $\beta$ denotes the average contemporaneous treatment effect, $\Gamma_i$ captures time-invariant unit heterogeneity, and $\Theta_t$ captures common shocks affecting all units in a given period. 
The error term $\epsilon_{it}$ collects residual variation and is allowed to exhibit arbitrary temporal dependence.

Under each assignment scheme considered, identification of $\beta$ follows from randomization
conditional on unit and time fixed effects. 
Our focus is therefore not on identification, but on how alternative assignment schemes affect the sampling variability of the TWFE estimator $\hat{\beta}$ when outcomes are serially correlated.

In the case where treatment is assigned at the unit level, identification of $\beta$ follows directly from the experimental design. In the pre-treatment period, no units receive treatment, so $D_{it}=0$ for all $i$. In the treatment period, treatment is randomly assigned to half of the units. Consequently, $\beta$ is identified from cross-sectional differences between treated and untreated units within the treatment period, after controlling for unit and time fixed effects.

The TWFE estimator serves two roles in our analysis. 
First, it delivers a consistent estimate of the common estimand $\beta$ under all designs. Second, and central to our contribution, it provides a unified basis for comparing how different treatment assignment schemes translate into differences in statistical precision. By holding the estimand, estimator, and outcome model fixed, we attribute differences in variance, false-positive rates, and effective runtime entirely to the design-induced variation in treatment timing. Subsequent sections build on this framework to analyze these effects theoretically, through simulation, and in an empirical application.

\subsection{Assignment Schemes: Fixed Assignment and SBs}
\label{sec:assignment_schemes}

We consider two classes of treatment assignment schemes that differ in how treatment status varies over time within units: fixed assignment and SB designs. Throughout, treatment status is encoded by the binary indicator $D_{it}$ in \eqref{eq:twfe}.

\paragraph{Fixed assignment.}
Under a fixed assignment scheme, each unit is assigned permanently to either treatment or control for the duration of the experiment. Formally, treatment status satisfies
\[
D_{it} = D_i \quad \text{for all } t,
\]
where $D_i \in \{0,1\}$ is fixed over time.
As a result, treatment varies only across units and not within units.
This design corresponds to the classical A/B testing paradigm commonly used in panel-based experiments.

\paragraph{SB designs.}
Under a SB design, treatment status varies within units over time.
Each unit alternates between treatment and control across pre-specified time blocks, so that
\[
D_{it} \in \{0,1\} \quad \text{and varies with } t \text{ for a given } i.
\]
The experiment horizon is partitioned into contiguous blocks of equal length, and treatment assignments are defined at the block level.
Within each block, treatment status is held fixed, while across blocks units switch between treatment and control.

SB designs are typically constructed to satisfy balance constraints, such as ensuring that each unit spends approximately the same number of blocks in treatment and control.
In addition, assignments may be coordinated across units to preserve balance at higher levels of aggregation, depending on operational requirements.

\paragraph{Comparison.}
The key distinction between the two schemes is whether treatment variation occurs within units over time.
Fixed assignment eliminates within-unit treatment variation by construction, whereas SB designs deliberately introduce such variation.
Holding the estimand and estimator fixed, this difference in assignment structure has direct implications for the sources of variation used in estimation and, consequently, for statistical precision.
These implications are analyzed in the next section.

\subsection{Inference and Standard Errors}

Inference is conducted without imposing independence or identical distribution assumptions on the error term $\epsilon_{it}$. To obtain standard errors that are comparable across assignment schemes, we use cluster-robust variance estimators that allow for arbitrary dependence within clusters (\cite{Arellano1987}, \cite{CameronMiller2015}).

Clustering is required because the sampling variability of $\hat{\beta}$ depends not only on the estimator, but also on the assumed dependence structure of the errors. By computing clustered standard errors consistently across all designs, we ensure that differences in estimated precision reflect differences in the information content generated by the assignment schemes, rather than artifacts of misspecified error assumptions.

The same clustering strategy is applied throughout the paper when evaluating precision under the null, false-positive rates, and power. Holding both the estimator and the variance estimation method fixed allows us to isolate the role of experimental design in determining the sampling variability of $\hat{\beta}$.

\subsection{Why SBs Increase Precision}
\label{sec:why_SBs}

This section explains why the SB assignment schemes introduced in Section~\ref{sec:assignment_schemes}
can yield substantially higher statistical precision than fixed assignment designs.
Throughout, we take the two-way fixed effects model in \eqref{eq:twfe} as given and focus on how alternative
assignment structures affect the sampling variability of $\hat{\beta}$, rather than its interpretation.

\paragraph{Within-unit comparisons and heterogeneous timing as sources of identifying variation.}
The identifying variation in a two-way fixed-effects (TWFE) model can be understood by examining how treatment variation survives successive removal of unit and time fixed effects. Applying the within-unit (demeaning) transformation removes all time-invariant unit-specific heterogeneity and yields
\begin{equation}
\tilde Y_{it}
=
\beta \bigl(D_{it} - \bar D_{i\cdot}\bigr)
+
\bigl(\Theta_t - \bar \Theta_{\cdot}\bigr)
+
\bigl(\epsilon_{it} - \bar \epsilon_{i\cdot}\bigr),
\label{eq:within_unit}
\end{equation}
where $\bar D_{i\cdot}$ denotes the unit-specific time average. Equation \eqref{eq:within_unit} shows that, after removing unit fixed effects, identification of $\beta$ relies on deviations of a unit’s treatment status from its own average. Units whose treatment status never changes over time therefore contribute no identifying variation after this transformation.

However, within-unit variation alone is not sufficient. If all units follow the same treatment timing, then the remaining variation in $D_{it} - \bar D_{i\cdot}$ is common across units and is absorbed by time fixed effects. To make explicit which treatment variation survives both unit and time fixed effects, define the twice-demeaned (two-way residualized) treatment indicator
\[
D^{**}_{it}
\equiv
D_{it}
-
\bar D_{i\cdot}
-
\bar D_{\cdot t}
+
\bar D_{\cdot\cdot},
\]
and analogously $Y^{**}_{it}$. The TWFE model can then be written as
\begin{equation}
Y^{**}_{it}
=
\beta D^{**}_{it}
+
\epsilon^{**}_{it},
\label{eq:twfe_double_demeaned}
\end{equation}
which makes clear that estimation of $\beta$ relies solely on the residualized treatment variation $D^{**}_{it}$.

Heterogeneous treatment timing across units ensures that $D^{**}_{it}$ is nonzero in periods where some units are treated while others are not. In such periods, treatment deviations are neither unit-specific constants nor common time shocks, and therefore survive both fixed effects. Repeated switches generate multiple such deviations, creating repeated within-unit treated-versus-control comparisons that accumulate identifying information. Under standard regularity conditions, the variance of the TWFE estimator satisfies
\[
\operatorname{Var}(\hat\beta \mid D)
\;\propto\;
\frac{1}{\sum_{i,t} (D^{**}_{it})^2},
\]
so that heterogeneous timing and repeated switches increase precision by increasing the effective variance of the residualized treatment regressor. By comparison, perfectly synchronized treatment timing implies $D^{**}_{it}=0$ for all $(i,t)$, in which case $\beta$ is not identified.

\paragraph{SBs reduce the impact of temporal dependence.}
In many operational settings, the error process $\epsilon_{it}$ in \eqref{eq:twfe} exhibits temporal dependence,
often in the form of positive autocorrelation. To capture this, we decompose
\begin{equation}
\epsilon_{it} = \eta_{it} + u_{it},
\label{eq:error_decomp}
\end{equation}
where $\eta_{it}$ denotes a persistent component and $u_{it}$ is weakly dependent noise.
Persistence here refers to slow decay of autocorrelation rather than nonstationarity.

Any treatment--control comparison ultimately reduces to a difference in outcomes, either across units at a given point in time or within the same unit across time. 
Under fixed assignment, treated and control units are distinct, so identification relies on cross-sectional treatment--control comparisons. 
Persistent components of the error process then enter these comparisons in levels, rather than through changes over time. 
As a result, slowly moving shocks $\eta_{it}$ do not cancel and directly inflate the variance of the treatment effect estimator. 
To see this, consider a cross-sectional comparison at time $t$,
\[
Y_{it} - Y_{jt},
\]
where unit $i$ is treated and unit $j$ is untreated. Using equation \eqref{eq:twfe},
this comparison can be written as
\[
Y_{it} - Y_{jt}
=
\beta
+
(\Gamma_i - \Gamma_j)
+
(\eta_{it} - \eta_{jt})
+
(u_{it} - u_{jt}),
\]
where the time fixed effect $\Theta_t$ cancels by construction. 
Both unit fixed effects and persistent shocks therefore enter the comparison in levels and do not cancel.

By contrast, under SB designs identification relies on within-unit comparisons across time. 
After accounting for unit and time fixed effects, the treatment effect is identified from changes in outcomes for the same unit as treatment status switches. 
For a given unit $i$, the relevant comparison takes the form
\[
Y_{i,t+h} - Y_{it}.
\]
Substituting the outcome equation yields
\[
\begin{aligned}
Y_{i,t+h} - Y_{it}
&=
\beta (D_{i,t+h}-D_{it})
+
(\Theta_{t+h}-\Theta_t) \\
&\quad
+
(\eta_{i,t+h} - \eta_{it})
+
(u_{i,t+h} - u_{it}),
\end{aligned}
\]
where the unit fixed effect $\Gamma_i$ cancels by construction. 
After accounting for time fixed effects, the persistent component enters the estimator only through differences over time, $\Delta_h \eta_{it} \equiv \eta_{i,t+h}-\eta_{it}$.

The variance of this difference is
\[
\mathbb{V}(\Delta_h \eta_{it})
=
2\gamma(0)-2\gamma(h),
\]
where $\gamma(h)=\mathrm{Cov}(\eta_{it},\eta_{i,t-h})$. When shocks are persistent, $\gamma(h)$ remains close to
$\gamma(0)$ for small $h$, implying that $\mathbb{V}(\Delta_h \eta_{it})$ is small. Consequently, slowly moving
components contribute much less to treated--control comparisons under SB designs. When treatment
alternates sufficiently frequently relative to the persistence of $\eta_{it}$, low-frequency shocks cancel
more effectively, reducing their impact on the variance of $\hat{\beta}$ and improving statistical
precision.

\paragraph{Effective sample size intuition.}
The effect of temporal dependence on precision can be summarized through the variance of a time average.
For a covariance-stationary process with autocovariances
$\gamma_k \equiv \mathrm{Cov}(\epsilon_{it},\epsilon_{i,t-k})$ and autocorrelations
$\rho_k \equiv \gamma_k/\gamma_0$, the sample mean
$\bar{\epsilon}_{i\cdot} \equiv T^{-1}\sum_{t=1}^T \epsilon_{it}$ satisfies
\begin{equation}
\mathbb{V}\!\left(\bar{\epsilon}_{i\cdot}\right)
=
\frac{\gamma_0}{T}
\left(
1 + 2 \sum_{k=1}^{T-1}
\left(1-\frac{k}{T}\right)\rho_k
\right),
\label{eq:ess}
\end{equation}
which follows from
$\mathbb{V}(\bar{\epsilon}_{i\cdot}) = T^{-2}\sum_{t=1}^T\sum_{s=1}^T \mathrm{Cov}(\epsilon_{it},\epsilon_{is})$
under stationarity (see, e.g., \cite{shumwaystoffer2017}).
Positive autocorrelation inflates $\mathbb{V}(\bar{\epsilon}_{i\cdot})$ relative to the i.i.d.\ case,
reducing the effective number of independent observations.

SB designs mitigate this inflation by constructing within-unit treated--control comparisons over
shorter horizons. When treatment alternates frequently and is balanced over calendar time, persistent
low-frequency components contribute less to these comparisons. As a result, SBs increase the
effective sample size and improve statistical precision.

\paragraph{Design implications.}
The precision gains from SB designs depend on the interaction between block length,
the persistence of the error process, and the presence of carryover effects.
Shorter blocks strengthen the reduction of persistent components but increase the risk of carryover,
while longer blocks reduce carryover at the cost of weaker variance reduction.
These trade-offs motivate the placebo analyses, type I error evaluations,
and power calculations presented in subsequent sections.

\section{Experimental Evaluation}

We evaluate how much temporal SB designs improve precision in LATAM’s seat-ancillary pricing experiments by comparing fixed pods vs weekly/daily SBs while holding the data and estimator constant. Using both a calibrated synthetic leg–day generator and a 2024 route–day panel (80 routes) with balanced k-means pods, we run placebos and injected uplifts to measure standard-error reductions, power, and Type I error over realistic experiment durations.

\subsection{Experimental Setup}

\subsubsection{Synthetic Data}
\label{sec:synth}
To evaluate temporal SB designs under dependence structures that are known to govern estimator precision, we construct a calibrated synthetic data generator based on LATAM's seat-ancillary outcomes aggregated to one observation per (leg, day). Recent analyses of SB experiments emphasize that performance is driven by temporal dependence and periodic structure, and by system-wide time variation that induces correlation across units \cite{bojinov2023design,hu2022SB,xiong2024data}. In our application, treatment-induced carryover across days is not expected to be material for seat-ancillary purchases: travelers primarily decide whether to purchase after conditioning on the fare price, and they typically observe the ancillary price (e.g., seat selection) only late in the checkout funnel. Consequently, we do not include an explicit carryover mechanism in the data-generating process. Instead, the generator focuses on reproducing the empirical temporal and cross-sectional structure most relevant for SBs in this setting: heterogeneous leg baselines, day-of-week effects, smooth longer-horizon seasonality, shared common shocks, and within-leg autocorrelation.

\textit{Generative model.}
For each leg $i$ and day $t$, we generate
\begin{equation}
y_{i,t}
=
\mu_i
+
\alpha_{d(t)}
+
\gamma_i \,\mathrm{Seas}_t
+
\beta_i \, S_t
+
r_{i,t},
\label{eq:data_generation}
\end{equation}
where $d(t)\in\{0,\dots,6\}$ is the day-of-week, $\alpha_{d(t)}$ is a mean-centered weekly profile, $\mathrm{Seas}_t$ is a smooth seasonal component, $S_t$ is a common time shock, and $r_{i,t}$ is a leg-specific autoregressive residual:
\begin{equation}
r_{i,t} = \phi_i r_{i,t-1} + \varepsilon_{i,t}, \qquad
\varepsilon_{i,t} \sim \mathcal{N}(0,\sigma_i^2).
\end{equation}

\textit{Fitting procedure (fit--remove--fit).}
A central issue is variance double-counting: if per-leg dispersion is fit on raw series and we later add seasonality/shocks/autocorrelation, the synthetic series becomes overly variable. We therefore estimate the components we inject, remove them, and fit dispersion on the remaining innovations:
\begin{enumerate}
    \item \textit{Baselines and day-of-week.} Estimate $\mu_i=\mathbb{E}[y_{i,t}]$ and a global weekly profile $\alpha_d$ from $y_{i,t}-\mu_i$, mean-centering $\alpha_d$.
    \item \textit{Smooth seasonality.} Fit $\mathrm{Seas}_t$ on the cross-leg daily average residual using a low-order Fourier basis \cite{hyndman2021fpp3}. Concretely, we parameterize a smooth seasonal component with $K$ harmonics and period $P$ as
    \begin{equation}
    \mathrm{Seas}_t
    =
    \sum_{k=1}^{K}
    \left[
    a_k \sin\!\left(\frac{2\pi k t}{P}\right)
    +
    b_k \cos\!\left(\frac{2\pi k t}{P}\right)
    \right],
    \label{eq:fourier_seasonality}
    \end{equation}
    where $K$ controls the complexity (larger $K$ allows higher-frequency seasonal variation) and $P$ is set to the seasonal period (e.g., $P \approx 365.25$ for annual seasonality on daily data).

    \item \textit{Common shocks.} After removing $\mu_i$, $\alpha_{d(t)}$, and $\gamma_i\mathrm{Seas}_t$, we form the residual matrix $R \in \mathbb{R}^{T \times N}$ (days $\times$ legs) and extract its dominant rank-1 component via PCA \cite{jolliffe2002pca}. We interpret the first principal-component score as a common time factor $S_t$ (standardized to unit variance) and the associated loadings as leg-specific exposures $\beta_i$, consistent with the use of principal components to estimate common factors in large panels \cite{stockwatson2002}.

    \item \textit{Autocorrelation and innovations.} After removing $\mu_i$, $\alpha_{d(t)}$, $\gamma_i\mathrm{Seas}_t$, and $\beta_i S_t$, we model the remaining within-leg dependence with an AR(1) residual process (a standard representation of within-series temporal dependence \cite{box2015tsa}). We estimate $\phi_i$ from the residual series for each leg and compute $\sigma_i$ from one-step prediction errors $\varepsilon_{i,t}=r_{i,t}-\phi_i r_{i,t-1}$.

\end{enumerate}

\textit{Cross-leg distributional modeling.}
To simulate new portfolios of legs, we model the cross-leg distribution of baselines and innovation scales:
(i) we fit a Gaussian mixture to $\log \mu_i$ (capturing multi-modality), and
(ii) we fit a log-linear conditional model
\begin{equation}
\log \sigma_i = a + b \log \mu_i + \eta_i, \qquad \eta_i \sim \mathcal{N}(0,\tau^2),
\end{equation}
which encodes empirical mean--variance scaling (Taylor's law) and prevents sampling $\mu_i$ and $\sigma_i$ independently \cite{taylor1961aggregation,perry1981taylorslaw}.

\textit{Controllable perturbations.}
The generator exposes interpretable multipliers to stress-test SB precision:
\emph{seasonality strength} ($\mathrm{Seas}_t \leftarrow c_{\mathrm{Seas}}\mathrm{Seas}_t$),
\emph{common-shock strength} ($S_t \leftarrow c_{\mathrm{Shock}}S_t$),
and \emph{AR persistence} (a stability-preserving scaling of $\phi_i$ that increases/decreases $|\phi_i|$ while remaining within the stationary region). 

\textbf{Synthetic Datasets for Evaluation.}
\label{sec:synth_datasets}
We construct four synthetic datasets to evaluate temporal SB estimators under an in-distribution baseline and three targeted departures that strengthen specific temporal components. All datasets are generated with the same calibrated simulator (Section~\ref{sec:synth}) using $N=80$ legs and a daily calendar; they differ only by scaling by the factor of 2 the magnitude of the corresponding component in Equation \ref{eq:data_generation}:

\begin{itemize}
  \item \textit{Baseline} uses fitted parameters without additional scaling and is designed to mimic the observed data-generating process.
  \item \textit{More seasonality} increases the magnitude of the long-horizon seasonal component , yielding larger peak-to-trough variation over longer horizons.
  \item \textit{More shocks} increases the magnitude of the shared common-shock component, yielding stronger cross-leg co-movement and higher aggregate volatility.

  \item \textit{More AR(1)} increases the strength of the leg-level AR(1) residual component, yielding greater day-to-day persistence in idiosyncratic deviations after accounting for common shocks and seasonal structure.
\end{itemize}

These departures represent plausible regime shifts relative to the historical calibration window. For example, long-horizon seasonality can be substantially stronger in markets with a higher share of leisure travel, while common shocks can intensify due to world events, network-wide operational disruptions, or policy and regulatory changes that affect ancillaries broadly (e.g., changes to baggage rules or fee disclosure requirements). Residual persistence AR(1) can also increase in practice, even after accounting for day-of-week patterns and smooth seasonality, due to inventory and booking dynamics and capacity and scheduling constraints. 

\textbf{Treatment Effect in Synthetic Data.}
Synthetic uplift of magnitude $\delta$ is imposed on treated observations by transforming the outcome metric:
\[
Y_{it}^{(\delta)} =
\begin{cases}
Y_{it}\,\big(1 + \delta + \eta_{it}\big), & \text{if } D_{it}=1, \\[4pt]
Y_{it}, & \text{otherwise},
\end{cases}
\]
where $\eta_{it} \sim N(0,\,\delta^2)$ is a mean-zero Gaussian shock whose variance is proportional to the level of $Y_{it}$. In this formulation, $\delta$ represents the average percentage uplift applied to treated route--day observations, while the stochastic term $\eta_{it}$ preserves the heteroskedasticity and multiplicative noise structure characteristic of ancillary revenue data. This procedure mirrors realistic improvements in ancillary monetization while maintaining the empirical distribution and temporal dependence present in the real world data.

\begin{table*}[t]
\centering
\caption{Type I Error and power on synthetic datasets.}
\label{tab:type1_error_power_summary}
\begin{tabular}{llrrrrrr}
\toprule
& & \multicolumn{2}{c}{2 weeks} & \multicolumn{2}{c}{8 weeks} & \multicolumn{2}{c}{16 weeks} \\
\cmidrule(lr){3-4}\cmidrule(lr){5-6}\cmidrule(lr){7-8}
Dataset & Design & Type I Error (\%) & Power (\%) & Type I Error (\%) & Power (\%) & Type I Error (\%) & Power (\%) \\
\midrule

\multirow{3}{*}{Baseline}
 & Fixed Pods        & 5.2 & 15.4 & 5.2 & 35.8 & 4.1 & 42.8 \\
 & Weekly SB & 6.5 & 27.3 & 5.3 & 74.0 & 6.2 & 95.2 \\
 & Daily SB  & 8.2 & 34.7 & 6.1 & 83.9 & 5.0 & 96.5 \\
\midrule

\multirow{3}{*}{More shocks}
 & Fixed Pods        & 5.4 & 14.9 & 5.3 & 33.5 & 3.8 & 36.0 \\
 & Weekly SB & 6.5 & 25.0 & 5.7 & 70.4 & 6.8 & 92.5 \\
 & Daily SB  & 8.2 & 32.4 & 6.8 & 80.9 & 5.6 & 95.0 \\
\midrule

\multirow{3}{*}{More seasonality}
 & Fixed Pods        & 5.0 & 15.1 & 5.1 & 29.0 & 4.2 & 30.2 \\
 & Weekly SB & 6.5 & 27.4 & 5.4 & 74.1 & 6.3 & 95.0 \\
 & Daily SB  & 8.2 & 35.0 & 6.1 & 84.0 & 5.2 & 96.2 \\
\midrule

\multirow{3}{*}{More AR(1)}
 & Fixed Pods        & 4.8 & 12.9 & 5.0 & 30.8 & 3.9 & 38.5 \\
 & Weekly SB & 5.8 & 23.2 & 5.3 & 66.9 & 6.3 & 91.8 \\
 & Daily SB  & 8.7 & 37.1 & 6.6 & 86.3 & 5.2 & 97.0 \\
\bottomrule
\end{tabular}
\end{table*}

\subsubsection{Real world data}

We use historical transactions from January 1 to December 31, 2024, covering long-haul and regional routes with paid seat selection. In LATAM’s network, long-haul routes are intercontinental flights linking South America with other continents, while regional routes operate within the same continent. 
All transactions are recorded at the booking-date level for economy-brand tickets, which exclude bundled ancillaries and therefore provide a clean baseline for measuring ancillary purchases.

The booking-date perspective plays a central role in our design. Because flights open for sale months in advance, booking windows overlap substantially across flights, inducing strong serial dependence across consecutive days. As a result, nearby dates often reflect similar underlying demand conditions, a feature that limits the amount of independent variation available for estimating temporal treatment effects.

The final dataset includes 80 non-directional routes used in the analysis. These routes differ substantially in their operational and commercial attributes—including baseline seat-take rate, ancillary revenue per passenger, distance, and ticket volume—capturing the cross-sectional heterogeneity that motivates much of the variance-reduction design in our experimental framework. We use a stratified randomization procedure that first clusters routes on standardized baseline characteristics (e.g., demand, monetization potential, and operational features) and then assigns routes within each cluster evenly into pods. The data construction and the randomization procedure are described in detail in the Appendix. %

\subsubsection{Evaluation Metrics}

For each assignment design we generate analyses over a grid of horizons ranging from 2 to 16 weeks. For a given horizon, we identify all feasible weekly start dates in 2024 that allow for symmetric pre- and post-periods, restricting attention to even durations to ensure clean alignment between treatment and control phases. For synthetic datasets we run both A/A tests and A/B tests with artificial uplift of 3\%; we repeat synthetic dataset generation procedure 20 times and for each dataset instance we create 10 different random splits into A and B pods. For the real dataset we run only A/A tests with 1000 different random splits. We report three complementary evaluation metrics:
\begin{itemize}
    \item \textit{Standard Error Reduction.} For each design and horizon, we compute the standard error (SE) of the estimated treatment effect under the corresponding analysis procedure. We report SE reduction relative to a fixed-pods baseline as
    $
        1 - \mathrm{SE}(\widehat{\tau}_{\text{design}})
        / \mathrm{SE}(\widehat{\tau}_{\text{fixed}}),
    $
    expressed as a percentage. This metric is directly interpretable as a proportional reduction in confidence interval width and reflects estimator efficiency.
    \item \textit{Type I Error.} Using A/A tests, we compute the empirical Type I error as the fraction of analyses in which the null hypothesis is rejected at the 5\% significance level (two-sided). Values close to the nominal 5\% indicate well-calibrated inference.
    \item \textit{Power.} Using A/B tests on synthetic datasets with a known multiplicative uplift of 3\%, we compute power as the fraction of analyses that reject the null hypothesis at the 5\% significance level (two-sided). Higher power indicates greater sensitivity to detect the effect at a fixed horizon and is reported alongside SE reductions to distinguish efficiency gains from potential miscalibration.
\end{itemize}

\subsection{Simulation Experiment}
\begin{table}[t]
\centering
\caption{SE reductions of SB designs on synthetic datasets.}
\label{tab:se_reduction_summary_pct}
\begin{tabular}{llrrrr}
\toprule
Dataset & SE reduction (\%) & \multicolumn{4}{c}{Duration (weeks)} \\
\cmidrule(lr){3-6}
 &  & 2 & 8 & 16 \\
\midrule

\multirow{3}{*}{Baseline}
 & Weekly SB vs Fixed Pods & 31.8 & 42.7 & 56.3 \\
 & Daily SB vs Fixed Pods  & 41.3 & 50.1 & 59.8 \\
 & Daily SB vs Weekly SB   & 13.9 & 13.0 & 8.0 \\
\midrule

\multirow{3}{*}{More shocks}
 & Weekly SB vs Fixed Pods & 30.2 & 44.5 & 63.0 \\
 & Daily SB vs Fixed Pods  & 40.5 & 52.5 & 66.7 \\
 & Daily SB vs Weekly SB   & 14.8 & 14.3 & 10.0 \\
\midrule

\multirow{2}{*}{More}
 & Weekly SB vs Fixed Pods & 32.6 & 54.4 & 68.6 \\
 & Daily SB vs Fixed Pods  & 42.1 & 60.6 & 71.2 \\
seasonality & Daily SB vs Weekly SB   & 14.1 & 13.4 & 8.5 \\
\midrule

\multirow{3}{*}{More AR(1)}
 & Weekly SB vs Fixed Pods & 33.1 & 43.7 & 55.4 \\
 & Daily SB vs Fixed Pods  & 50.1 & 57.3 & 64.2 \\
 & Daily SB vs Weekly SB   & 25.4 & 24.2 & 19.6 \\
\bottomrule
\end{tabular}
\end{table}

Table~\ref{tab:se_reduction_summary_pct} summarizes precision gains from SB designs relative to fixed pods and highlights how these gains depend on the dominant temporal structure. Across the calibrated baseline and the regimes with amplified common shocks or smooth seasonality, both weekly and daily SBs substantially reduce standard errors, with improvements increasing with duration (e.g., in the baseline, weekly: 31.8\% $\rightarrow$ 56.3\% and daily: 41.3\% $\rightarrow$ 59.8\% from 2 to 16 weeks). Moreover, strengthening common shocks or smooth seasonality generally increases the advantage of SBs over fixed pods, consistent with the intuition that more within-leg temporal variation can be differenced out more effectively under SB assignment than under a static pod partition (e.g., at 16 weeks, weekly improves from 56.3\% in the baseline to 63.0\% with more shocks and 68.6\% with more seasonality, while daily improves from 59.8\% to 66.7\% and 71.2\%, respectively). In these settings, the additional efficiency of daily over weekly switching remains present but modest (roughly 8.0\%--14.8\%), suggesting that both designs similarly benefit from averaging out shared and periodic variation over longer horizons. By contrast, under stronger residual persistence (``more AR(1)''), daily switching becomes materially more efficient than weekly switching, with the daily--weekly SE reduction rising to 25.4\% at 2 weeks and remaining 19.6\% at 16 weeks. This pattern aligns with the effect of serial correlation in reducing the effective sample size of temporally adjacent observations: more frequent alternation better neutralizes persistent idiosyncratic deviations that are not absorbed by common shocks or seasonality. Importantly, this is not merely a synthetic artifact; in applied work on additional datasets we have at times observed much larger separations between weekly and daily SB precision, suggesting that pronounced route-level idiosyncratic dynamics and short-memory persistence are common in real operational data and can meaningfully influence the relative performance of SB regimes.

Table~\ref{tab:type1_error_power_summary} reports Type I error and power for the three designs at selected durations. Type I error is estimated from A/A tests and power is estimated from A/B tests with a true treatment effect of 3\%. Across all scenarios, empirical Type I error is within statistical error of the nominal level given our Monte Carlo sample size of 200 runs, indicating that the procedures are well-calibrated. The only notable deviation occurs for the daily SB at the shortest duration (2 weeks), where rejection rates are somewhat elevated; a plausible explanation is that very short-horizon daily alternation is more sensitive to transient start-up effects and imperfect temporal balance (e.g., partial-week composition and residual serial dependence interacting with the assignment schedule), which can inflate finite-sample variance estimates and lead to over-rejection. In terms of power, both SB designs substantially outperform fixed pods in every regime, consistent with their sizable standard-error reductions. Daily SB achieves the 80\% power threshold by 8 weeks in all scenarios, whereas weekly SB reaches 80\% only at longer horizons, remaining below 80\% at 8 weeks (66.9\%--74.1\%). Fixed pods fail to reach 80\% power at any duration or in any scenario, with power remaining below 43\% even at 16 weeks. These results reinforce that more frequent alternation yields faster accumulation of effective information, and that SB designs can dramatically improve both time-to-detection and attainable power under realistic temporal dependence.

\subsection{Real Data Experiment}

\begin{table}[t]
\centering
\caption{SE reductions of SBs on the seats revenue dataset.}
\label{tab:real_se_reduction_summary_pct}
\begin{tabular}{lrrr}
\toprule
 SE reduction (\%) & 2 weeks & 8 weeks & 16 weeks \\
\midrule

  Weekly SB vs Fixed Pods & 32.5 & 50.2 & 67.7 \\
  Daily SB vs Fixed Pods  & 39.5 & 54.0 & 67.1 \\
  Daily SB vs Weekly SB & 10.4 & 7.7 & -1.9 \\
\bottomrule
\end{tabular}
\end{table}

\begin{table}[t]
\centering
\caption{Type I Error on the seats revenue dataset.}
\label{tab:real_type1_error_summary_pct}
\begin{tabular}{lrrr}
\toprule
 Design & 2 weeks (\%) & 8 weeks (\%) & 16 weeks (\%) \\
\midrule
 Fixed Pods        & 4.8 & 4.7 & 4.1 \\
 Weekly SB & 6.7 & 5.5 & 5.5 \\
 Daily SB  & 7.6 & 5.5 & 4.5 \\
\bottomrule
\end{tabular}
\end{table}

\begin{figure}
    \centering
    \captionsetup{justification=l, singlelinecheck=false, font=small}
    \captionsetup[subfigure]{justification=l, singlelinecheck=false, font=small}

    \begin{subfigure}[t]{0.49\linewidth}
        \centering
        \includegraphics[width=\linewidth]{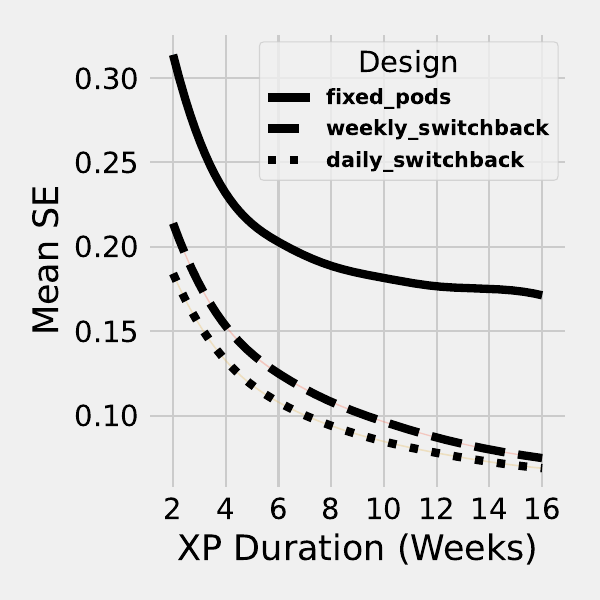}
        \caption{Calibrated synthetic data.}
        \label{fig:te_se_over_time_synth}
    \end{subfigure}
    \hfill
    \begin{subfigure}[t]{0.49\linewidth}
        \centering
        \includegraphics[width=\linewidth]{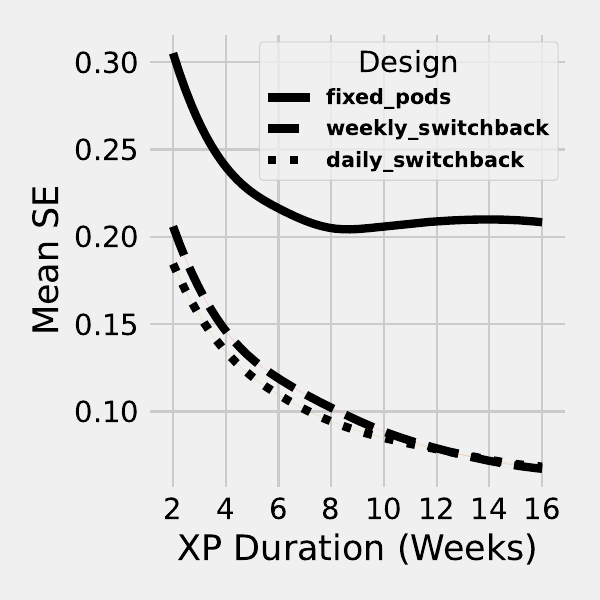}
        \caption{Seat pricing dataset.}
        \label{fig:te_se_over_time_real}
    \end{subfigure}

    \caption{Standard deviation of the estimated treatment effect across experimental horizons, by assignment scheme.}
    \label{fig:te_se_over_time}
\end{figure}

Tables~\ref{tab:real_se_reduction_summary_pct} and~\ref{tab:real_type1_error_summary_pct} report results on the real-world dataset and broadly corroborate the main conclusions from the synthetic evaluation. First, SB designs materially improve precision relative to fixed pods at all durations: weekly SB reduces SE by 32.5\%, 50.2\%, and 67.7\% at 2, 8, and 16 weeks, respectively, while daily SB achieves 39.5\%, 54.0\%, and 67.1\%. Consistent with the synthetic findings, the relative advantage of daily switching is most pronounced early in the experiment window: at 2 weeks, daily provides an additional 10.4\% SE reduction over weekly, and at 8 weeks the incremental gain remains positive (7.7\%). At 16 weeks, however, the daily--weekly difference slightly reverses (-1.9\%), even though both SB variants deliver very large gains over fixed pods. We view this late-horizon reversal as unlikely to reflect a true structural disadvantage of daily switching and more plausibly as a finite-sample artifact arising from the particular realization of calendar composition, residual dependence, and variance estimation at longer horizons, where both designs already achieve substantial variance reduction and the remaining differences are small. Finally, Type I error on the real dataset is generally close to nominal across designs and durations (Table~\ref{tab:real_type1_error_summary_pct}) and Figure \ref{fig:type_I_error_rate}, with mild over-rejection for daily SB  at the shortest duration that attenuates over time.

Together, these patterns indicate that SBs dramatically accelerate how quickly an experiment accumulates statistically informative variation. Baseline cross-sectional A/B tests remain noisy even after months of data, whereas SB designs reach high precision within a commercially realistic window.

\section{Discussion}
\subsection{Operational Implications}

Our results have direct implications for experimentation programs in operational environments where unit-level randomization is infeasible and outcomes are strongly time-dependent. First, across both synthetic regimes and the real-world airline dataset, SB designs consistently deliver substantial reductions in sampling variability relative to fixed pod assignment, translating into meaningfully faster time-to-decision for small effects. This improvement is operationally valuable precisely in the settings where traditional route- or geo-level A/B tests are most fragile: persistent cross-unit heterogeneity, common shocks, and calendar structure inflate uncertainty and can render long-running experiments impractical. Second, daily SBs tend to be especially beneficial early in the experiment window. In short horizons, more frequent alternation increases the number of within-leg contrasts and more effectively neutralizes transient idiosyncratic deviations, improving the stability of effect estimates and accelerating the point at which business-relevant power thresholds are achieved. Weekly SBs still provide large gains over fixed pods and may be easier to operationalize, but the empirical pattern suggests that teams seeking fast iteration can often justify the additional operational complexity of daily switching. 

\subsection{Limitations and Directions for Future Work}

A key limitation of our study is that we do not explicitly model \emph{temporal interference}---settings in which treatment at time $t$ affects outcomes at later times beyond what is captured by standard serial dependence. Examples include intertemporal substitution and ``shopping'' behavior (e.g., users delaying or accelerating purchases in response to anticipated prices), as well as spillovers that propagate through shared capacity, inventory, or congestion over time. In our application this concern is limited because passengers see prices of ancillaries at the end of the booking flow reducing scope for meaningful intertemporal substitution in the outcome we study. However, in many operational experiments temporal interference is central, and both SB estimands and optimal cadences can change when carryover effects are present. An important direction for future work is therefore to extend our evaluation framework to settings with explicit carryover and spillover dynamics, and to assess design and analysis strategies---such as washout periods, staggered switching, or estimators that incorporate lagged treatment exposure---that preserve valid inference under temporal interference.

\section{Conclusion}

We evaluate temporal SB designs for experiments in operational settings where unit-level randomization is infeasible and outcomes exhibit strong temporal structure, including calendar effects, common shocks, and serial dependence. Across calibrated synthetic regimes and real-world data, SBs substantially improve precision relative to fixed pod assignment, translating into faster time-to-detection for small effects, with daily switching providing the largest benefits early in the experiment window. These results offer actionable guidance for designing temporally structured experiments under practical constraints and support SBs as an effective default for accelerating reliable decision-making.

\bibliographystyle{ACM-Reference-Format}
\bibliography{sample-base}

\appendix

\begin{figure}[b]
    \centering
    \includegraphics[width=0.99\linewidth]{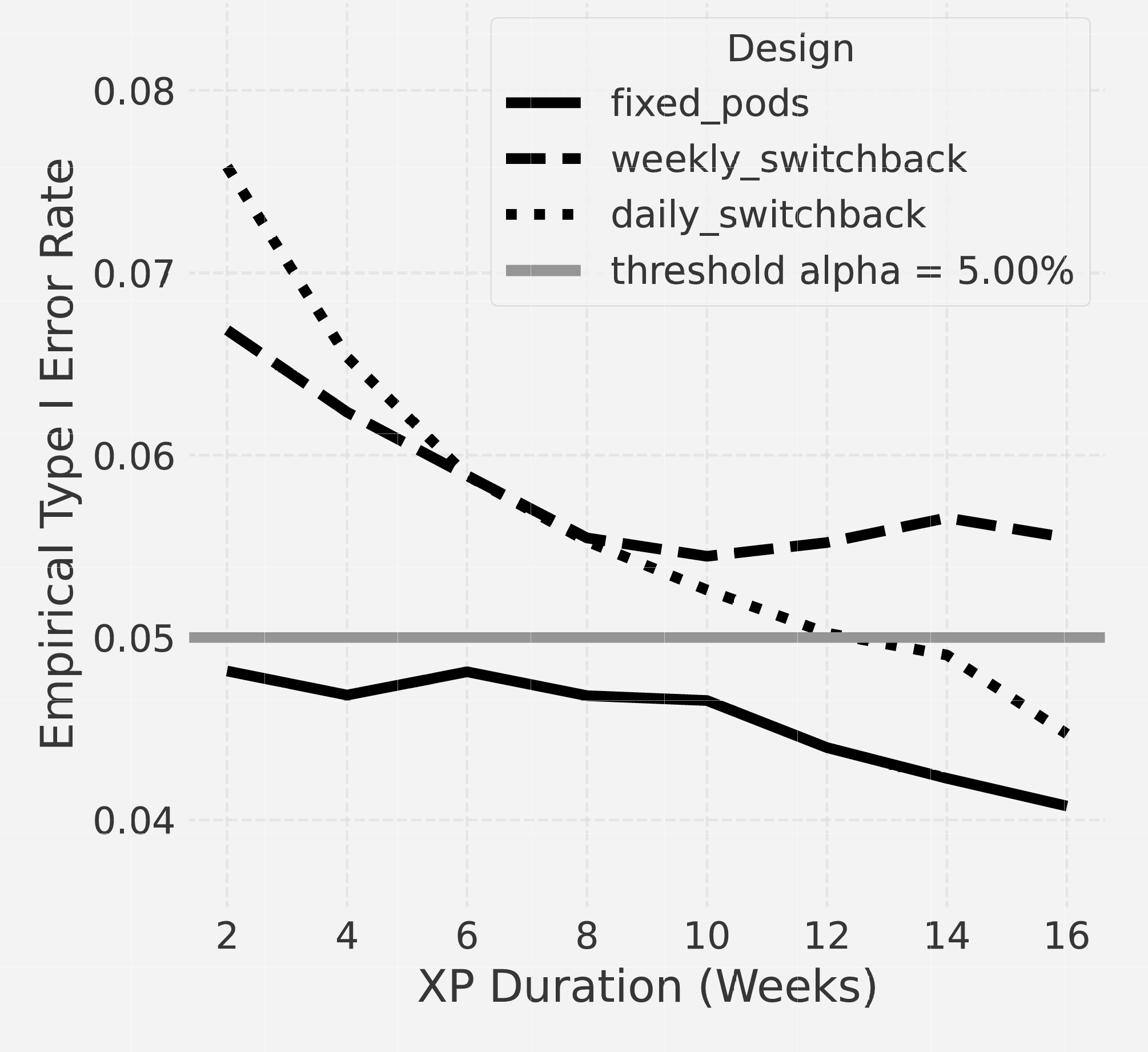}
    \captionsetup{justification=l, singlelinecheck=false, font=small}
    \caption{Empirical type~I error rate across experimental horizons, by assignment scheme. The horizontal line marks the nominal 5\% significance level.}
    \label{fig:type_I_error_rate}
\end{figure}

\section{Empirical Experiment Details}

From these records, we construct a directional route--day panel dataset (treating A$\rightarrow$B and B$\rightarrow$A as distinct units) containing daily aggregates of key metrics such as tickets sold, seat-take rate, and seat revenue per ticket. This level of aggregation aligns with the structure of LATAM’s pricing systems, where seat-product fare rules are defined for directional origin–destination pairs.

In real experiments, to ensure consistent treatment exposure for round-trip passengers, we apply treatment symmetrically across both directions of a route; therefore we adhere to this rule in this study as well. Although the dataset is organized directionally, the effective unit of treatment assignment is therefore the \textit{non-directional route--day}, meaning that both directions of a city pair share the same treatment status on any given date.

\textit{Balanced Assignment via Clustering and Pods.}
To ensure balanced treatment assignment, we construct pods based on route-level characteristics using a $k$-means clustering algorithm. Because routes differ widely in their baseline demand, ancillary monetization potential, and operational characteristics, balancing treatment and control groups requires grouping routes that are broadly comparable along these dimensions. We first standardize features such as seat-take rate, average seat revenue per passenger, distance, and average tickets sold, and then apply $k$-means to partition routes into clusters whose members are similar in these attributes. Within each cluster, routes are randomly assigned to Pod A or Pod B in roughly equal proportions.

This stratified procedure preserves the broad heterogeneity across clusters while improving covariate balance within experimental groups, consistent with the recommendations of Morgan \& Rubin (2012) and Rubin \& Imbens (2015). To incorporate sampling variability in the cluster-formation process and avoid dependence on a single realization, we repeat the clustering and randomization 1{,}000 times, generating a library of balanced yet distinct route partitions. This library allows us to assess the robustness of our conclusions across alternative, but comparably balanced, experimental allocations and to ensure that our findings do not hinge on any particular random assignment.

\end{document}